\documentclass[a4paper,11pt]{article}
\pdfoutput=1
\usepackage{jcappub}
\usepackage{mathtools}
\usepackage{ulem}
\usepackage{bm}
\usepackage{xcolor}

\renewcommand{\thesection}{\arabic{section}}

\def\laq{~\raise 0.4ex\hbox{$<$}\kern -0.8em\lower 0.62ex\hbox{$\sim$}~}
\def\gaq{~\raise 0.4ex\hbox{$>$}\kern -0.7em\lower 0.62ex\hbox{$\sim$}~}

\def\beq{\begin{equation}}
\def\eeq{\end{equation}}
\def\bea{\begin{eqnarray}}
\def\eea{\end{eqnarray}}
\def\bean{\begin{eqnarray*}}
\def\eean{\end{eqnarray*}}

\def\laq{~\raise 0.4ex\hbox{$<$}\kern -0.8em\lower 0.62ex\hbox{$\sim$}~}
\def\gaq{~\raise 0.4ex\hbox{$>$}\kern -0.7em\lower 0.62ex\hbox{$\sim$}~}

\def\be{\begin{equation}}
\def\ee{\end{equation}}
\def \ga {\gamma}

\def\beq{\begin{equation}}
\def\eeq{\end{equation}}
\def\bea{\begin{eqnarray}}
\def\eea{\end{eqnarray}}

\def \pa {\partial}
\def \ra {\rightarrow}
\def \ti {\widetilde}
\def \La {\Lambda}

\newcommand{\Ups}{\Upsilon}

\newcommand{\Scal}{\mathcal S}

\newcommand{\Hcal}{\mathcal H}
\newcommand{\Ical}{\mathcal I}
\newcommand{\Pcal}{\mathcal P}
\newcommand{\Mcal}{\mathcal M}

\def\laq{~\raise 0.4ex\hbox{$<$}\kern -0.8em\lower 0.62ex\hbox{$\sim$}~}
\def\gaq{~\raise 0.4ex\hbox{$>$}\kern -0.7em\lower 0.62ex\hbox{$\sim$}~}

\def\beq{\begin{equation}}
\def\eeq{\end{equation}}
\def\bea{\begin{eqnarray}}
\def\eea{\end{eqnarray}}
\def\bean{\begin{eqnarray*}}
\def\eean{\end{eqnarray*}}

\def \pa {\partial}
\def \ra {\rightarrow}
\def \ti {\widetilde}

\def \La {\Lambda}
\def \Da {\Delta}
\def \da {\delta}
\def \b {\beta}
\def \a {\alpha}

\def \ga {\gamma}

\def \Sg {\Sigma}
\def \da {\delta}

\def \Om {\Omega}

\def \U{\Upsilon}

\title{Generalized covariant prescriptions for averaging cosmological observables}

\author[a,b]{G. Fanizza,}
\author[c]{M. Gasperini,}
\author[d]{G. Marozzi,}
\author[e]{G. Veneziano}

\affiliation[a]{Istituto Nazionale di Fisica Nucleare, Sezione di Pisa, Italy}
\affiliation[b]{Instituto de Astrofis\'ica e Ci\^encias do Espa\c{c}o,
Faculdade de Ci\^encias da Universidade de Lisboa,
Edificio C8, Campo Grande, P-1740-016, Lisbon, Portugal}
\affiliation[c]{
Dipartimento di Fisica, Universit\`a di Bari, 
Via G. Amendola 173, 70126 Bari, Italy,\\
and Istituto Nazionale di Fisica Nucleare, Sezione di Bari, Italy
}
\affiliation[d]{
Dipartimento di Fisica, Universit\`a di Pisa, Largo B. Pontecorvo 3, 56127 Pisa, Italy,\\
and Istituto Nazionale di Fisica Nucleare, Sezione di Pisa, Italy
}
\affiliation[e]{CERN, Theory  Department, CH-1211 Geneva 23, Switzerland,\\
and Coll\`ege de France, 11 Place M. Berthelot, 75005 Paris, France\\}

\emailAdd{gfanizza@fc.ul.pt}
\emailAdd{gasperini@ba.infn.it}
\emailAdd{giovanni.marozzi@unipi.it}
\emailAdd{gabriele.veneziano@cern.ch}

\abstract{We present two new covariant and general prescriptions for averaging scalar observables on spatial regions typical of the observed sources and intersecting the past light-cone of a given observer. One of these prescriptions is adapted to sources exactly located on a given space-like hypersurface, the other applies instead to situations where the physical location of the sources is characterized by the experimental ``spread" of a given observational variable.
The geometrical and physical differences between the two procedures are illustrated by computing the averaged energy flux received by distant sources located on (or between) constant redshift surfaces, and by working in the context of a perturbed $\La$CDM geometry. We find significant numerical differences (of about ten percent or more, in a large range of redshift) even limiting our model to scalar metric perturbations, and stopping our computations to the leading non-trivial perturbative order.}

\keywords{Light-cone average, observational cosmology, luminosity-redshift relation, cosmological backreaction
 
\vskip18pt 

\noindent{\bfseries\large\sffamily{Preprints:}} BA-TH/721-19, CERN-TH-2019-183
}

\begin{document}

\maketitle

\section{Introduction}
\label{Sec1}
\setcounter{equation}{0}

The choice of a correct procedure for averaging physical observables in a curved space-time is not only an important formal problem for any geometric theory of gravity, but also a crucial ingredient of observational cosmology. 

For instance, the possible impact of small scale inhomogeneities on the large scale dynamics cannot be properly addressed without using a well-posed prescription for averaging their contribution to the cosmological equations. In addition, recent results in the context of numerical relativity have stressed the need for a full theoretical control on the choice of the averaging procedure \cite{Adamek:2018rru}. Starting with the right choice is crucial for reaching the sought level of precision (or, more ambitiously, for writing the correct numerical code) in the context of modern cosmological simulations \cite{Adamek:2015eda,Giblin:2015vwq,Bentivegna:2015flc,Macpherson:2018btl}.

In view of the many theoretical and phenomenological implications of these problems, several motivated proposals have been presented and discussed, during the last years, for averaging cosmological observables on both space-like and null (hyper)surfaces \cite{Buc1,Buc2,Gas1,Gas2,Marozzi:2010qz,1,Bonvin:2015kea,Yoo,Buchert:2018yhd,Buc3} (see also the reviews \cite{rev1,rev2} and references therein).  In such a context, for an unambiguous and well-posed prescription, various peculiar aspects of the problem have to be considered and clearly specified. For instance:

\begin{itemize}
\item{} \textit{Which} physical observables we are considering.
\item{} \textit{Who} is performing observations and in which state of  motion.
\item{} \textit{Which} type of messengers the observer is receiving.
\item{} \textit{Where} are  the sources located \textit{when} they emit the  messenger.
\end{itemize}
All the above ingredients provide indeed crucial contributions to the definition of the average integral and, in particular, to the specification of the  window function selecting the appropriate integration domain in the given space-time manifold. Leaving details to the following Sections, let us briefly introduce here the basic idea. 

Suppose, for instance, that we want to average an observable $\Scal$ which is measured through the light-like signals emitted by sources lying on a  space-like  hypersurface $\Sigma (A)$, defined by the condition $A(x)= A_0=$ const (where $A$ is a scalar field with time-like gradient). 
 Clearly, the light-like signals will originate from the (co-dimension 2) intersection of such hypersurface with the past light-cone of the observer, the latter being  specified by the value of a scalar field $V(x)=V_0$ with a light-like gradient. The corresponding average prescription is thus defined on a two-dimensional surface (if we are in four space-time dimensions), and is naturally characterized by an integration measure proportional to the proper area of the above-mentioned surface. The latter can  be written in general, for the intersection of two (or more) arbitrary hypersurfaces (see Eq. (2.10) of \cite{1}), in the form of a general-covariant integral over the space-time manifold 
${\Mcal_4}$, and reduces, in the case at hand, to \cite{1}:
\beq
 \int_{\Mcal_4}d^4x\sqrt{-g}\,\da (V_0-V)\da (A_0-A)\left|\pa_\mu V \pa^\mu A  \right|.
\label{11}
\eeq

As we will discuss in the next Section, this first kind of average can still take different explicit forms, and leads to the first class of general averaging prescriptions proposed in this paper. To be more explicit, let us give also a very simple example concerning the observation  of standard-candle sources. If we can measure their redshift we can then consider a free-falling observer receiving photons emitted from  sources that are located on constant-redshift surfaces. On the other hand, if we can also measure the angular size of those sources, we can relate their luminosity to their angular sizes. In that case we can still consider a free-falling observer, receiving photons, however, from sources that now lie on constant angular-distance surfaces. When averaging the corresponding observational data we find that the two setups lead to different window functions selecting different integration domains, and thus corresponding to different averaging integrals. In principle, there is also the possibility of receiving different messengers from the same source: the fact that such signals may travel along different paths \cite{Fanizza:2015gdn} may lead, again, to different window functions, different average integrals and thus different averaged results, even if the properties of the source and of the observer are the same.

There is however a different direction in which we can generalize the above prescription, and which leads to the second class of averaging procedures proposed in this paper. This is directly inspired by a close contact with the observational approach, and is motivated by a (possibly realistic) experimental situation where the effective location of the sources, differently from Eq. \eqref{11}, is not exactly specified by a given, geometrically well-defined hypersurface but is controlled instead by the unavoidable range of ``spread" of a physical observational variable. As a consequence,  the sources are in general confined within a thin space-time layer bounded by two very close hypersurfaces. The differences from the first average prescription also survive in the limit in which the data bin typical of the spread is very small, and the thickness of such layer tends to zero. This second prescription reproduces, in a particular case, the averaging procedure recently discussed in \cite{Yoo,2}.

To be more explicit let us also recall that in general, for a finite layer of thickness $\Delta$, the averaging prescription  was also essentially given in \cite{1} as:
\beq
 \int_{\Mcal_4}d^4x\sqrt{-g}\,\da (V_0-V)\Theta (A_0+\Delta-A)\Theta (A-A_0)\frac{\left|\pa_\mu V \pa^\mu A  \right|}{\sqrt{- \pa_\nu A \pa^\nu A}},
\label{finitelayer}
\eeq
 where, with respect to  Eq. (2.7) of \cite{1}, we have added a second Heaviside $\Theta$-function to restrict the integration to the layer. One could naively expect that, by going to the $\Delta \rightarrow 0$ limit, Eq. \eqref{finitelayer} would go smoothly over to Eq. \eqref{11} but this turns out to be incorrect: the product of the two $\Theta$-functions does go the Dirac $\delta$-function of \eqref{11}, but a non-trivial extra weight factor remains (as will be explicitly shown in Sect. \ref{Sec2}). Essentially, this means that the infinitesimal width of the layer is non necessarily constant all along $\Sigma(A)$, effect that is lost if we go directly to the zero-width limit.

We shall apply both prescriptions to averaging the distance-redshift relation in a perturbed cosmological background, considering in particular the effects of scalar metric perturbations on the radiation flux received from distant astrophysical sources. In that case it will be shown that the two prescriptions give the same results only when limiting ourselves to contributions arising from the second radial derivatives of the velocity potential (more precisely, from the so-called effect of ``redshift space distortion"), while  there are differences already to the first perturbative order when considering all leading contributions (including, in particular, the so-called Doppler terms). This clearly demonstrates the physical difference between the two prescriptions.

This paper is organized as follows. In Sect. \ref{Sec2} we define the two new averaging prescriptions. In Sect. \ref{Sec3} we specialize them to the case of constant redshift hypersurfaces, and we explicitly write their expressions using for the metric the Geodesic Light-Cone (GLC) gauge \cite{1}. In Sect. \ref{Sec4} we apply the two averaging prescriptions to a cosmological geometry including scalar perturbations to the leading non-trivial order.  We explicitly compute the average and the fractional corrections of the radiation flux received from distant astrophysical sources as a function of their redshift $z$, taking into account all the leading order effects such as Doppler, lensing, and redshift space distortion. In Sect. \ref{Sec5} we present a further example illustrating the possible role of non-geometric weight factors included into the integral measure of the averaging prescriptions. Sect. \ref{Sec6} is devoted to our conclusive remarks. In the Appendix \ref{app:A} we finally provide the technical details  needed for the numerical evaluation of the leading contributions to the average integrals.


\section{General prescriptions for light-cone averaging}
\label{Sec2}
\setcounter{equation}{0}

In this paper we are mainly interested in defining covariant average procedures that are relevant for cosmological observations based on light-like signals. To this purpose we need to specify the following main ingredients.

\begin{itemize}
\item{} A scalar field $\Scal(x)$ whose average we are interested in.
\item{} A scalar field $\rho(x)$ which specifies an additional weight factor associated with the averaging of the
variable $\Scal(x)$ (such as, for instance, the total matter density).
\item{}A scalar field $A(x)$, with timelike gradient, often conveniently associated with a chosen free-falling observer whose four-velocity is given by $n_\mu= -\pa_\mu A /|\pa_\nu A \pa^\nu A |^{1/2}$.
\item{}A scalar field $V(x)$, with lightlike gradient\footnote{In the case of massive messenger, of course, we should consider a field $V$ with timelike gradient.}, that identifies the past light-cones centered on the observer, and spanned by the null momenta $k_\mu= \pa_\mu V$ of the photons emitted by the  sources ($k_\mu k^\mu=0$).
\item{} A scalar field $B(x)$ which identifies the space-like (hyper)surfaces on which the sources are located. 
\item{}Finally, a scalar field $C(x)$ whose normalized gradient $m_\mu= -\pa_\mu C /|\pa_\nu C \pa^\nu C |^{1/2}$ defines, as better specified below, the flow lines along which we may consider the variation of the volume integral on the hypersurface identified by $B$ through the embedding higher-dimensional space-time. 
\end{itemize}

It should be noted that the choice of the scalar fields $B$ and $C$ is closely related to the geometrical background and to the type of (averaged) observations we are performing. We may be interested, for instance, in sources lying on constant-redshift spheres if we want to study the distance-redshift relation. In that case the natural choice is $B= k^\mu n_\mu$,  which specifies the redshift $z$ of the emitted photons as measured by the free-falling observer (we recall  that $1+z= (k^\mu n_\mu)/(k^\mu n_\mu)_o$, where $``o"$ denotes the observer position). The simplest physical situation suggests the choice $C=A$, corresponding to $m_\mu=n_\mu$. But other choices for $B$ and for $C$ are also possible if we are interested in different types of measurements and/or we are working in different physical contexts.

Given the above ingredients, we can now introduce a covariant prescription for averaging a physical (scalar) observable $\Scal$ on the two-dimensional spacelike region $\Sg (B_s)$, defined by the intersection of the source hypersurface $B=B_s$ with the given observer's past light-cone, $V=V_o$. Starting with the covariant four-volume integral, and following the same procedure already illustrated in \cite{1} (but with a more general window function), we then define the average
\beq
\langle \Scal \rangle_{\Sg (B_s)} =\frac{I(\Scal, \rho,V_o, A, B_s, C)}{I(1, \rho,V_o, A, B_s, C)}, 
\label{21}
\eeq
where 
\bea
\!\!\!\!\!\!
I(\Scal,\rho, V_o, A, B_s, C) &=& \int_{\Mcal_4}d^4x\,\sqrt{-g}\, \Scal \rho \,n^\mu\nabla_\mu\Theta(V_o-V)
\,m^\mu\nabla_\mu\Theta(B_s-B)
\nonumber \\ &=&
 \int_{\Mcal_4}d^4x\,\sqrt{-g}\,\Scal  \rho \,\da (V_o-V)\da (B_s-B)\,{\pa^\mu A \pa_\mu V\over
 \left|\pa_\a A \pa^\a A \right|^{1/2}} \,{\pa^\nu C \pa_\nu B\over
 \left|\pa_\b C \pa^\b C \right|^{1/2}},
 \nonumber \\ &&
 \label{22}
 \eea
and where we have used the properties of the Heaviside step function $\Theta$ and of the Dirac $\delta$-function.

Note that for $\rho=1$ and $A=B=C$ one exactly recovers the averaging prescription adopted in \cite{1} (see Sect. \ref{Sec3}), which is covariant and also invariant under the general reparametrization $A \ra \ti A(A,V)$ and $V \ra \ti V(A,V)$. The generalized prescription (\ref{22}), on the contrary, is covariant but invariant only under separate reparametrization of the different scalar fields, $A \ra \ti A(A)$,  $B \ra \ti B(B)$,  $C \ra \ti C(C)$ and  $V \ra \ti V(V)$. We shall consider and discuss possible physical choices of $C$ and $B$ in the following section. 

Let us now consider a second (and different) covariant  averaging prescription, motivated by a -- possibly more realistic -- experimental situation where the physical location of the sources is not exactly specified by the hypersurface $B=B_s$, but is characterized by a ``spread"  of the variable $B$ within a bin $\Da B_s$, with $\Da B_s \ll \mathcal{R}(B)$, where $\mathcal{R}(B)$ is the size of the whole range of $B$. In that case we are led to define a new average for our observable $\Scal$ as
\beq
\langle \Scal \rangle_{\Da B_s} =\frac{J(\Scal, \rho,V_o, A, B_s, \Da B_s)}{J(1,\rho, V_o, A, B_s, \Da B_s)}, 
\label{23}
\eeq
where
\beq
J(\Scal,\rho, V_o, A, B_s, \Da B_s)= \int_{\Mcal_4}d^4x\,\sqrt{-g}\, \Scal \rho\,n^\mu\nabla_\mu\Theta(V_o-V)\Theta(B_s+ \Da B_s-B) \Theta(B-B_s).
\label{24}
\eeq

With such a new window function we are limiting the integration  volume to a region corresponding to a finite range of the scalar field $B$, namely to   $B_s < B< B_s + \Da B_s$. For  $\Da B_s \ll \mathcal{R}(B)$, in particular, we can expand the step function $\Theta(B_s+ \Da B_s -B)$ and we obtain, in the limit $\Da B_s \ra 0$,
\beq
\Theta(B-B_s)\Theta(B_s+ \Da B_s-B) \simeq  \Da B_s \,\da (B_s - B) +{\cal{O}}(\Da B_s^2).
\label{25} 
\eeq
The average integral (\ref{24}) thus reduces to
\beq
J(\Scal, \rho, V_o, A, B_s, \Da B_s)= \Da B_s \int_{\Mcal_4}d^4x\,
\sqrt{-g}\, \Scal \rho\,\da (V_o-V) \da (B_s-B)\,{\pa^\mu A \pa_\mu V\over
\left|\pa_\a A \pa^\a A \right|^{1/2}}\,. 
\label{26}
\eeq
Obviously, the constant factor $\Da B_s$ drops out in the ratio defining the averaging prescription (\ref{23}), and  we are lead to a final surface integral defined on the intersection between the light-cone and the hypersurface $B=B_s$, exactly as before. As before, the integral (\ref{26}) is covariant and separately invariant under the scalar reparametrizations $A \ra \ti A(A)$,  $B \ra \ti B(B)$ and  $V \ra \ti V(V)$. However, the surface  integration of Eq. (\ref{26}) is weighted by a factor which is different  in general from that of Eq. (\ref{22}), and the two averaging prescriptions  (\ref{21}), (\ref{23}) may coincide, in general, only if the  expression
\beq
\pa^\nu C \pa_\nu B/\left|\pa_\mu C \pa^\mu C\right|^{1/2}
\label{27}
\eeq
factorizes out of the integrals, and thus simplifies in the ratio defining the averaging prescription (\ref{21}).

Some physical differences between the two averages (\ref{21}) and (\ref{23}) will be illustrated in the  following sections. We shall first concentrate on the geometric ingredients of the average integrals  putting everywhere $\rho=1$, and we will discuss some possible  interpretations of the scalar fields $B$ and $C$ working in the context of the convenient Geodesic Light-Cone (GLC) gauge \cite{1} (see also \cite{Fleury2} for a pedagogical introduction to the GLC coordinates). An example of averages including a non-trivial scalar field $\rho(x)$ will be finally illustrated in Sect. \ref{Sec5} of this paper.

\section{Averages on constant-redshift surfaces in the GLC gauge}
\label{Sec3}
\setcounter{equation}{0}

From now on we shall consider sources localized on or between constant-redshift surfaces, $z=z_s$ (with a possible spread controlled by a redshift bin $\Da z \ll z$). Hence, 
we have to select a field $B$ which can be  directly associated with
the redshift $z$ of the observed sources. 

In such a  context we can conveniently use the so-called GLC coordinates $x^\mu= (\tau, w, \ti \theta^a)$, $a=1,2$, where the most general cosmological metric can be parametrized in terms of the six arbitrary function $\U$, $U^a$, $\ga_{ab}= \ga_{ba}$, and the line element takes the form \cite{1}
\beq
ds^2_{GLC}=-2\U dwd\tau+\U^2dw^2+\ga_{ab}\left(d\ti\theta^a-U^a dw\right)\left(d\ti\theta^b-U^b dw\right).
\label{31}
\eeq
The corresponding inverse metric $g^{\mu\nu}_{GLC}$ (that we report here for later use) is given by
\beq
g^{\mu\nu}_{GLC} =
\left(
\begin{array}{ccc}
-1 & -\Upsilon^{-1} & -U^b/\Upsilon \\
-\Upsilon^{-1} & 0 & \vec{0} \\
-(U^a)^T/ \Upsilon & \vec{0}^{\, T} & \gamma^{ab}
\end{array}
\right).
\label{32}
\eeq
We recall that $w$ is a null coordinate, that photons travel along geodesics with constant $w$ and $\tilde\theta^a$, and that $\tau$ coincides with the time coordinate of the synchronous gauge \cite{BenDayan:2012pp}. In the GLC gauge we can thus perform averages defined on the past light-cone of a free-falling observer, according to the prescriptions of Sect. \ref{Sec2}, by simply identifying \cite{1} $A=\tau$ and $V=w$.

In that case we obtain $n_\mu= -\da_\mu^\tau$, $k_\mu= \pa_\mu w$ and (using the metric \ref{31}) $n^\mu k_\mu= \U^{-1}$. It follows that the redshift $z$ of a signal received at the time $\tau_o$, and traveling along the light-cone $w=w_o$, is controlled by the ratio
\beq
1+z = {\U_o\over \U},
\label{33}
\eeq
where $\U_o= \U(\tau_o,w_o,\ti \theta^a)$ and  $\U= \U(\tau,w_o,\ti \theta^a)$. In the so-called ``temporal gauge" of the GLC coordinates \cite{Fleury2}, where $\tau_o=w_o$ and  $\U_o=1$, we can thus relate the field $B$ to the redshift parameter $z$ simply by choosing $B=\U^{-1}$ (similarly, when dealing with observational angles, it would be useful to further specify the GLC gauge according to Ref. \cite{Fanizza:2018tzp}). Finally, by computing the determinant of the metric (\ref{31}), we obtain 
$\sqrt{-g}=\U \sqrt{\ga}$ where $\ga= {\rm det} \,\ga_{ab},$ and we can rewrite the integral prescription (\ref{22}) as follows:
\beq
I(\Scal, w_o, \tau,z_s,C)
=\int_{\Mcal_4}d\tau dw \,d^2 \ti\theta \, \sqrt{\ga}\,\Scal\,\delta(w_o-w)\delta(z_s-z)\,
m^\nu \pa_\nu \U^{-1}
\label{34}
\eeq
where we have set $\rho=1$, as anticipated. The vector field $m^\mu(C)$ is left unspecified for the moment.

The integration on $d\tau$, on the other hand, can be transformed into an integral over the redshift variable by using Eq. (\ref{33}), which gives (recalling that both $w$ and $\ti\theta^a$ are constant along the relevant null geodesics)
\beq
d\tau=-\frac{\U^2}{\pa_\tau\U}dz.
\label{35}
\eeq
Eq. (\ref{34}) thus reduces to
\beq
I(\Scal, w_o, z_s,C)
=\int_{\Sg_s}d^2 \ti\theta \left[ \sqrt{\ga}\,\Scal\,
\frac{m^\nu \pa_\nu \U}{\pa_\tau \U}\right]_{w_o,z_s},
\label{36}
\eeq
where $\Sg_s$ is the two-dimensional surface determined by the intersection of the past light-cone $w=w_o$ with the redshift sphere $z=z_s$, and all the integrated functions are to be evaluated at $w=w_o$, $z=z_s$.

We have still to specify $C$, in order to explicitly compute the vector field $m_\mu= \pa_\mu C /|\pa_\nu C \pa^\nu C |^{1/2}$. Let us consider here two motivated possibilities. 

\begin{itemize}
\item{}A first possibility is $C=A=\tau$. In that case sources and observer evolve through the embedding spacetime along flow lines generated by the same (timelike) tangent vector field, $m^\mu =n^\mu \equiv - g^{\mu\tau}_{GLC}$. Using the metric (\ref{32}) the integral (\ref{36}) thus becomes
\beq
I(\Scal, w_o, z_s,\tau)
=\int_{\Sg_s}d^2 \ti\theta \left[ \sqrt{\ga}\,\Scal\,
\left( 1+\frac{1}{\U}\frac{\pa_w \U}{\pa_\tau \U}+\frac{U^a}{\U}\frac{\pa_a \U}{\pa_\tau \U}\right)\right]_{w_o,z_s}.
\label{37}
\eeq

\item{}A second possibility is $C=B=1+z$. In that case the flow lines describing the evolution of the constant-redshift hypersurfaces are generated by the gradients of the redshift field itself, i.e. $m^\mu= \pa^\mu \U^{-1}/ \left|\pa_\nu \U^{-1} \pa^\nu \U^{-1} \right|^{1/2}$. The integral (\ref{36}) becomes
\beq
I(\Scal, w_o, z_s,z)
=\int_{\Sg_s}d^2 \ti\theta \left[ \sqrt{\ga}\,\Scal\,
\frac{\left|g^{\mu\nu}_{GLC}\,\pa_\mu \U\pa_\nu\U\right|^{1/2}}{\pa_\tau \U}\right]_{w_o,z_s},
\label{38}
\eeq
and, using the metric (\ref{32}), it can be explicitly rewritten as
\beq
I(\Scal, w_o, z_s,z)
=\int_{\Sg_s}d^2 \ti\theta \left[ \sqrt{\ga}\,\Scal\,
\left| 1+\frac{2}{\U}\frac{\pa_w \U}{\pa_\tau \U}+\frac{2U^a}{\U}\frac{\pa_a \U}{\pa_\tau \U}-\ga^{ab}\frac{\pa_a\U\pa_b\U}{(\pa_\tau\U)^2}
\right|^{1/2}\right]_{w_o,z_s}.
\label{39}
\eeq
\end{itemize} 

Clearly, the two averages corresponding to Eqs. (\ref{37}) and (\ref{39}) are in general different at the level of exact integral prescriptions; however, they both give the same result for a perturbed cosmological metric, at the first perturbative order. In fact, by expanding the small perturbations of the cosmological geometry around the zeroth-order (homogeneous, isotropic) background, one finds non-vanishing contributions to $\pa_w\U$, $\pa_a\U$ and $U^a$ only by including perturbations to linear (or higher) order (see Sect. \ref{Sec4}); on the contrary, $\pa_\tau \U$ is non-vanishing already on the background (see e.g. \cite{1} for the explicit expression of the FLRW metric in GLC coordinates). Hence,  Eqs. (\ref{37}) and (\ref{39}) lead, to first order, to the same approximate  integral (see also Sect. \ref{Sec4}): 
\beq
\int_{\Sg_s}d^2 \ti\theta \left[ \sqrt{\ga}\,\Scal\left(1+\frac{1}{\U}\frac{\pa_w\U}{\pa_\tau\U}\right)\right]_{w_o,z_s} + \cdots
\label{310}
\eeq

It may be important to note, at this point, that if we are working at the first perturbative order then the average integral of  Eq. (\ref{22}) 
is always independent on the field $C$, for any possible choice of of the scalar fields $A$, $B$, $C$ specifying our averaging prescription. In fact, starting with the general form of Eq. (\ref{22}) (with $\rho=1$), and expanding as before the  geometry described by the metric (\ref{31}), we obtain, to first order, 
\bea
\!\!\!\!\!\!
I(\Scal,\rho, V_o, A, B_s, C) &=& \int_{\Mcal_4}d^4x\,\sqrt{-g}\, \Scal  \,n^\mu\nabla_\mu\Theta(V_o-V)
\,m^\mu\nabla_\mu\Theta(B_s-B)
\nonumber \\ &=&
\int_{\Sg_s}d^2 \ti\theta \left[ \sqrt{\ga}\,\Scal\left(1-\frac{1}{\U}\frac{\pa_w A}{\pa_\tau A}+\frac{1}{\U}\frac{\pa_w B}{\pa_\tau B}\right)\right]_{w_o,B_s} + \cdots
\label{Gen310}
 \eea
where $\Sigma_S$ is now the two-dimensional surface where the given scalar field $B$ takes constant values. Such a first-order result holds quite independently of the choice of the scalar field $C$. Eq.(\ref{310}), in particular, is immediately recovered by identifying $A$ with $\tau$ and $B$ with the redshift parameter.

It is also interesting to compare the above results in Eqs. (\ref{310}) and (\ref{Gen310}) with the much simpler surface integral
\beq
\int_{\Sg_s}d^2 \ti\theta \left(\sqrt{\ga}\,\Scal\right)_{w_o,z_s},
\label{311}
\eeq
obtained in the context of a similar prescription for light-cone averages, proposed in \cite{1} and studied in previous papers \cite{BenDayan:2012pp,BenDayan2,BenDayan3,BenDayan:2013gc,3}.
The result (\ref{311}) can be exactly reproduced (even if $B$ is not identified with the redshift parameter) 
within the more general approach of this paper (i.e., starting from Eq. (\ref{22})) by choosing $\rho=1$, $V=w$ and $A=B=C$.  
Indeed, in that case, none of the additional terms depending on the gradients of $\U$ (and present in both Eqs. (\ref{37}) and (\ref{39})) can be generated, and Eq. (\ref{22})  immediately leads to the pure (and invariant under general reparametrizations) surface integral (\ref{311}). 

\begin{figure}[t!]
\centering
\includegraphics{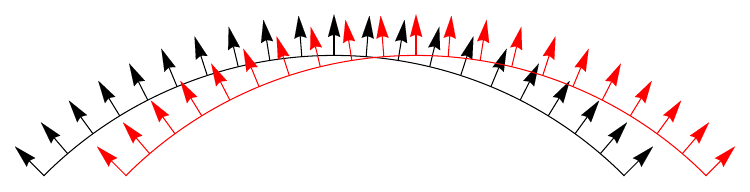}
\caption{We consider a constant-time  (black curve) and a constant-redshift (red curve) hypersurface. The arrows represent the respective variation fields at a given time or redshift. When specifying an averaging prescription, we have to choose from which hypersurface we are starting, and along which field we are moving. The case $B=1+z$, $C=A=\tau$ discussed in this section (see Eq. (\ref{37})) refers to constant-redshift hypersurfaces with a flow driven by time gradients (red curve and black arrows, not shown in the picture). The case $C=B=1+z$ (see Eq. (\ref{39})) refers instead to constant-redshift hypersurfaces connected by redshift gradients (red curve and red arrows). The black curve with black arrows, on the contrary, represents constant-time hypersurfaces connected by time gradients, i.e. $C=B=\tau$. 
}
\label{f1}
\end{figure}

Note that the Eq. (\ref{311}) represents an exact, non-perturbative result  
once one assumes $A=B=C$. On the contrary, in order to recover the same result at the first perturbative order, the choice $A=B$ is already enough (see Eq. (\ref{Gen310})). See also Fig. \ref{f1} for a simple graphical illustration of different possible choices of the averaging scalar fields $A$, $B$ and $C$.

In order to conclude this section, let us also present the explicit form assumed in the GLC gauge by the integral (\ref{26}), defining the light-cone average (\ref{23}) for sources characterized by an observational bin $\Da B_s$.

When applying the prescription (\ref{26}) there is no ambiguity due to the choice of the $C$ field, and we can follow exactly the same procedure adopted for the integral (\ref{22}). We thus identify $V=w$, $A=\tau$, $B= \U^{-1}$, and the integral (\ref{26}) (with $\rho=1$) becomes
\beq
J(\Scal, w_o, z_s,\Delta z_s)
= - \Da z_s \int_{\Sg_s}d^2 \ti\theta \left[ \sqrt{\ga}\,\Scal\,
\frac{\U^2}{\pa_\tau \U}\right]_{w_o,z_s}.
\label{312}
\eeq
Again the result in Eq. \eqref{312} is different in general from the former prescription \eqref{311}, and different as well from the generalized prescriptions \eqref{37} and \eqref{39}. 

To make contact with previous papers let us note that the above result \eqref{312}, with the weight $\rho$ included and identified with the density $\rho_s$ of the sources, may exactly coincide (in an appropriate limit) with the so-called {\it number-count average} used in \cite{Yoo,2}. Consider in particular the following integral measure \cite{2}:
\beq
{\rho_s \Delta z\, d^2_A\over (1+z) H_{||}} \,d\Om,
\label{1}
\eeq
where all quantities are evaluated on the past light-cone and at constant redshift $z_s$ (see Eqs. (2.11) and (2.12) of \cite{2}). Here $\Delta z$ is the (small) finite bin of redshift data, $\rho_s$ is the related volume density of sources, $d_A$ their angular distance and $d\Om$ the corresponding angular integration measure. Finally, $H_{||}$ is the local longitudinal expansion parameter defined in general by $H_{||}= (1+z)^{-2} k^\mu k^\nu \nabla_\mu u_\nu$, where $k_\mu$ is the (usual) photon momentum, and $u_\mu$ the local velocity of the matter sources.

Let us move now to the coordinates of the GLC gauge, where \cite{BenDayan:2013gc}  $d_A^2 \,d\Om= \sqrt{\ga}\, d^2 \ti\theta$ (see also \cite{DiDio:2014lka} for an explicit expression of the  number-count in the GLC metric). Also, let us consider the particular case in which the velocity field $u_\mu$
appearing in the definition of  $H_{||}$ may be chosen to be the same as (or proportional to) the velocity $n_\mu$ of our class of free-falling observers. In that case, and in the GLC gauge where $u_\mu=n_\mu= -\da_\mu^\tau$ and $k^\mu= g^{\mu w}$ we then obtain, using Eq. (\ref{33}):
\beq
H_{||}= - {1\over \Ups_o^2} {\pa_\tau \Ups\over \Ups}.
\label{2}
\eeq
Using as before the temporal gauge $\Ups_o=1$, the expression (\ref{1}) thus reduces to
\beq
- d^2 \ti\theta
\left[ \Da z \, \rho_s \sqrt{\ga} \,{\Ups^2\over \pa_\tau \Ups} \right]_{w_o,z_s},
\label{3} 
\eeq
which clearly coincides with our averaging prescription (\ref{312}), for any given observable $\Scal$, provided we include the additional weight factor $\rho_s$ (see Sect. \ref{Sec5} for an explicit numerical example).

It should be stressed, finally,  that all the new averages based on the integrals (\ref{37}), (\ref{39}) and (\ref{312}) may coincide with the old prescription of Ref. \cite{1}, based on Eq. (\ref{311}), only if we are working in a homogeneous and isotropic FLRW metric background, but for a more general perturbed geometry they are all different, in principle,  already at the first perturbative order.  Possible observable consequences of the differences among the various averaging prescriptions will be illustrated in the following sections.

\section{Comparing different averaging prescriptions}
\label{Sec4}
\setcounter{equation}{0}

In this section we will compare the averaging prescriptions based on the integrals (\ref{310}), (\ref{311}) and (\ref{312})  in a cosmological geometry which includes scalar metric perturbations. 
As will be explained below (see in particular the discussion following Eq. (\ref{49})), for the computations to be performed in this paper, concerning the geometric contributions to the integration measure appearing in the various averaging prescriptions, it will be enough to limit ourselves to the first perturbative order. Assuming the absence of anisotropic stresses we can parametrize the scalar perturbations with a single Bardeen potential $\psi$, so that the linearly perturbed metric in the Poisson gauge, using polar angles $(\theta, \phi)$ and conformal time $\eta$, takes the form
\beq
ds^2_{PG}=a^2(\eta)\left[ -\left( 1+2\,\psi \right)d\eta^2+(1-2\,\psi)\left( dr^2 +r^2d\theta^2+r^2\sin^2\theta d\phi^2\right) \right].
\label{41}
\eeq

For an explicit computation of the average integrals (\ref{310})--(\ref{312}) we need to express the perturbed geometry (\ref{41}) in the GLG gauge. To this purpose, following \cite{3}, it is convenient to introduce the coordinate system $y^\mu=\left( \eta,\eta^+,\theta,\phi \right)$, where 
$\eta^+=r+\eta$, so that the metric (\ref{41}) becomes
\bea
ds^2_{PG}=g_{\mu\nu}^{PG}dy^\mu dy^\nu
&\equiv&a^2(\eta)\left[ -4\psi\,d\eta^2
+( 1-2\psi)( d{\eta^+}^2-2d\eta d\eta^+ )\right.\nonumber\\
&&\left.
-\left( 1-2\psi \right)( \eta^+-\eta)^2\left( d\theta^2+\sin^2\theta d\phi^2 \right) \right].
\label{42}
\eea
Considering the coordinate transformation $ x^\mu \ra y^\mu(x)$ 
(where $x^\mu=( \tau,w,\tilde\theta^a )$ are GLC coordinates) we have
\beq
g^{\mu\nu}_{PG}(y)=\frac{\pa y^\mu}{\pa x^\alpha}\frac{\pa y^\nu}{\pa x^\beta}\,g^{\alpha\beta}_{GLC}(x),
\label{43}
\eeq 
where $g^{\alpha\beta}_{GLC}$ is the metric (\ref{32}), while $g^{\mu\nu}_{PG}$ is the inverse of the metric tensor (\ref{42}).

We have to compute, in particular, the three different integration measures appearing in Eqs. (\ref{310})--(\ref{312}), including in the geometry (expressed in GLC form) all contributions arising from the Bardeen potential $\psi$, up to first order. 
Following the procedure (and the results) of previous papers (see in particular \cite{3}, where similar computations have been performed by consistently including all second order perturbative contributions) we thus expand the coordinate transformation as $y^\mu(x)=y^{\mu}_{(0)}+y^{\mu}_{(1)} + \cdots$, and linearize the perturbed GLC metric by defining  $\Upsilon=\Upsilon^{(0)}+\Upsilon^{(1)}$, $U^a=U^{a}_{(0)}+U^a_{(1)}$,  $\ga_{ab}=\gamma_{ab}^{(0)}+\gamma_{ab}^{(1)}$. 
The (unperturbed) background quantities are given by (see e.g. \cite{1,3}):
\bea
&&
\eta^{(0)}(\tau)=\int_{\tau_{in}}^\tau \frac{d\tau'}{a(\tau')},~~~~~\eta^{+\,(0)}=w,~~~~~ \theta^{a}_{(0)}=\tilde\theta^a,
\nonumber\\ &&
\Upsilon^{(0)}=a(\tau), ~~~~~~~~~~~~~~~
U^{a}_{(0)}=0, ~~~~~~~~~\ga_{ab}^{(0)}=a^2 r(\tau,w)^2\text{diag}( 1,\sin^2\tilde\theta^1).
\label{44}
\eea
Here $r(\tau,w)=w-\eta^{(0)}(\tau)$, and $\tau_{in}$ corresponds to an early enough time when perturbations were negligible.

The integral measure (\ref{311}), in particular, is completely specified by the element of proper area $d^2 \mu=d^2 \ti \theta |\det \ga_{ab}|^{1/2}$, whose explicit perturbed expression has already been computed in \cite{BenDayan:2012pp,
BenDayan2,BenDayan3,BenDayan:2013gc}. Hence, for the new averaging prescriptions of this paper, we only need to take into account the  corrections to the above measure as they appear under the two integrals (\ref{310}) and (\ref{312}). 

By exploiting the results of a detailed computations of the various components of Eq. (\ref{43}), presented in \cite{3}, we obtain in particular that the measure correction of Eq. (\ref{310}) can be written to first order as follows:
\bea
\left(1+{1\over \U}{\pa_w \U\over \pa_\tau \U}\right)_{w_o,z_s} &=& 1- v_{\rVert s}
+ \frac{1}{\Hcal_s}\left[\pa_rv_{\rVert s}+\pa_r\psi_s+2\pa_\eta\psi_s
+2\int_{\eta_s}^{\eta_o}d\eta\,\pa_\eta^2\psi\left( \eta,\eta_o-\eta,\theta^a \right)\right]\nonumber\\
&&+\mathcal{O}(\psi^2) , 
\label{45}
\eea
where the subscript $s$ denotes that all the variables are evaluated at the source coordinates $\eta_s$, $r_s$.
Here $\Hcal=a'/a$ (the prime denotes differentiation with respect to $\eta$), and $v_{\rVert s}$ is the so-called velocity perturbation (or Doppler term), projected along the (unperturbed) radial direction connecting source and observer. This is given by
\beq
v_{\rVert s}= - \pa_r \eta^{(1)}_s, ~~~~~~~~~~~~
\eta^{(1)}_s=-\int_{\eta_{in}}^{\eta_s}\,d\eta\frac{a(\eta)}{a(\eta_s)}\psi(\eta,r_s,\theta^a).
\label{46}
\eeq
Following \cite{3} (see also \cite{Fanizza:2015gdn}) we have neglected in Eq. (\ref{45}) perturbative contributions from the peculiar velocity and from the gravitational (Bardeen) potential evaluated at the observer position. 
Indeed, the first type of terms can always be removed by going to the CMB frame. The second type of terms is important to regularize the formal infrared divergence of super-horizon fluctuations (as shown in \cite{Biern:2016kys} for the variance of the luminosity distance-redshift relation). In this work this problem is avoided by imposing a physical infrared cutoff at the horizon scale (see below, Eq. \eqref{423a}), which leaves us with negligible contributions at the observer positions.

Similarly, and with the same assumptions as before about the perturbative contributions evaluated at the observer position, the measure correction of Eq. (\ref{312}) can be written to first order as follows:
\bea
\left(\frac{\U^2}{\pa_\tau \U}\right)_{w_o,z_s}
&=&
\frac{a^2_s}{\Hcal_s}
\left[ 1+\psi_s
+\frac{1}{\Hcal_s}\left(\pa_r v_{\rVert s}+ \pa_\eta\psi_s \right) -
\right. \nonumber\\ &-&
\left.
\left( 1-\frac{\Hcal_s'}{\Hcal_s^2} \right)\left( v_{\rVert s}+\psi_s
+2\int_{\eta_s}^{\eta_o}d\eta\,\pa_\eta\psi\left( \eta,\eta_o-\eta, \theta^a \right) \right)
\right] + \mathcal{O}(\psi^2) .
\label{47}
\eea
This last result is in perfect agreement with the evaluation independently performed in \cite{DiDio:2014lka} with a different approach. Note that the homogeneous term ${a^2}/{\Hcal}$, multiplying the square brackets in the above equation, factorizes out of the integral (\ref{312})  and obviously drops out in the ratio (\ref{23}) defining the final averaging prescription. The physical differences from the previous measure (\ref{45}) are thus entirely due to the contribution of the first-order perturbations.

We are now in the position of discussing the physical differences among the various averaging procedures, induced by their different geometric ingredients.

\subsection{Example: fractional corrections to the flux average}
\label{Sec41}

The averaging prescription (\ref{311}) has been applied in previous papers \cite{BenDayan:2012pp,BenDayan3,BenDayan:2013gc} to estimate the geometric backreaction due to metric perturbations, arising in the computation of the averaged luminosity distance ${\langle d_L \rangle}(z)$. 
Working  with the associated observation variable, namely the received flux $\Phi(z)\sim d^{-2}_L(z)$, we have computed in previous papers \cite{BenDayan:2012pp,BenDayan:2013gc} the {\it ensemble} (or statistical) average (denoted by an overbar) of the geometric light-cone average (denoted by brackets) of the flux: namely, the quantity $\overline{\langle \Phi \rangle}$. 
Such results for the averaged flux may also represent a starting point for the computation of the averaged {\it flux drift} effect (see e.g. \cite{fluxdrift}), which we are planning to study in a future paper.

Let us recall, in this respect, that by working in a more general geometric context perturbed up to second order \cite{BenDayan:2012pp}, 
by expanding the flux variable as $\Phi \simeq \Phi^{FLRW}(1+\delta \Phi^{(1)}+\delta\Phi^{(2)}+ \cdots)$, and using the ``old" integral measure of Eq. (\ref{311}), expanded as $d^2\mu \simeq d^2 \mu^{(0)}(1+\da\mu^{(1)}+ \da\mu^{(2)}+  \cdots)$, the result for $\overline{\langle \Phi \rangle}$ can be written, to second perturbative  order, as follows
\beq
\overline{\langle \Phi \rangle}(z)=
\Phi^{FLRW}\left[1+f_\Phi(z)\right].
\label{48}
\eeq
Here $\Phi^{FLRW}$ is the unperturbed value of $\Phi$ computed in the FLRW metric background, and the corresponding fractional correction $f_\Phi(z)$ is given by \cite{BenDayan:2012pp}
\beq
f_\Phi(z) =
\overline{\langle \delta \Phi^{(2)} \rangle_{0}} 
+\overline{\langle \delta\mu^{(1)}\delta \Phi^{(1)} \rangle_{0}}
-\overline{\langle \delta\mu^{(1)}\rangle_{0}\langle \delta \Phi^{(1)} \rangle_{0}}\,\,\,,
\label{49}
\eeq
where $\langle \cdots \rangle_{0}$ denotes standard angular average performed with respect to the unperturbed measure 
$d^2 \mu^{(0)}$ of the FLRW geometry, and we have used the fact that {\it ensemble} averages do {\it not} factorize, i.e. $\overline{AB} \ne \overline{A} ~\overline{B}$.
As clearly stressed by the above result, it turns out that, even working at the second perturbative order, there are contributions to the fractional correction $f_\Phi$ from the second-order perturbations of the averaged variable, $ \delta \Phi^{(2)} $, but {\it not} of the integration measure \cite{BenDayan:2012pp} (namely, {\it no contributions} from $\da\mu^{(2)}$).  Hence, for the purpose of this paper of comparing the possible physical differences due to different definitions of the average integral, the perturbed results for the integration measures consistently computed up to first order, and reported in Eqs. (\ref{45}) and (\ref{47}), will be enough (as we have anticipated at the beginning of Sect. \ref{Sec4}).
See also Appendix \ref{app:A} for more details on the {\it ensemble} average procedure applied to a stochastic background of scalar perturbations. 

The above result for $f_\Phi(z)$, computed with the averaging prescription of Eq. (\ref{311}), has already been plotted in \cite{BenDayan3,BenDayan:2013gc} for a perturbed CDM and $\La$CDM cosmological geometry, including also the contributions of perturbations evaluated at the observer position\footnote{In this paper we will not include such contributions in the expression for $f_\Phi(z)$, in order to be consistent with the assumption made in Eqs. \eqref{45} and \eqref{47}.}. Let us now compute the same fractional correction, $\overline{\langle \Phi/ \Phi^{FLRW}\rangle}-1$, in the same geometry, using however for the light-cone average the two new prescriptions (\ref{21}) and (\ref{23}) proposed in this paper, and specified in particular by the integration measures of Eqs. (\ref{310}) and (\ref{312}). 

The perturbative expansion of the flux variable is the same as before, and the only difference is an additional, first-order contribution of the perturbed geometry to the generalized integration measures, which now can be expanded as follows: 
\beq
d^2\mu \simeq d^2 \mu^{(0)}(1+\da\mu^{(1)}+ \da m^{(1)}+ \cdots),
\label{410}
\eeq
where $\da\mu^{(1)}$ is the same term appearing in Eq. (\ref{49}),  arising from the perturbations of the measure (\ref{311}). 
The new terms $ \da m^{(1)}$, coming from the first-order perturbations of the modified measures, is given by our previous results (\ref{45}) and (\ref{47}). In particular, for the averaging prescription (\ref{21}) we have, from Eq. (\ref{45}):
\beq
\da m^{(1)}_{\Sg(B_s)}=
- v_{\rVert s}
+ \frac{1}{\Hcal_s}\left[\pa_rv_{\rVert s}+\pa_r\psi_s+2\pa_\eta\psi_s
+2\int_{\eta_s}^{\eta_o}d\eta\,\pa_\eta^2\psi\left( \eta,\eta_o-\eta,\theta^a \right)\right].
\label{411}
\eeq
For the averaging prescription (\ref{23}) we have, from Eq. (\ref{47}):
\beq
\da m^{(1)}_{\Da B_s}=
\psi_s
+\frac{1}{\Hcal_s}\left(\pa_r v_{\rVert s}+ \pa_\eta\psi_s \right) 
- \left( 1-\frac{\Hcal_s'}{\Hcal_s^2} \right)\left[ v_{\rVert s}+\psi_s
+2\int_{\eta_s}^{\eta_o}d\eta\,\pa_\eta\psi\left( \eta,\eta_o-\eta, \theta^a \right) \right].
\label{412}
\eeq
The new fractional corrections for the averaged flux variable, computed according to the standard procedure illustrated in \cite{BenDayan:2012pp}, but using the averaging prescriptions (\ref{310}), (\ref{312}) -- namely, using the generalized measure perturbations of Eq. (\ref{410}) -- can be finally expressed as follows:
\beq
\overline{\langle \Phi/ \Phi^{FLRW}\rangle}_{\Sg(B_s)}-1=
f_\Phi(z) +b_{\Sg(B_s)}(z),
\label{413}
\eeq
for the light-cone average (\ref{310}), and 
\beq
\overline{\langle \Phi/ \Phi^{FLRW}\rangle}_{\Da B_s}-1=
f_\Phi(z) +b_{\Da B_s}(z),
\label{414}
\eeq
for the light-cone average (\ref{312}). We have used for $f_\Phi(z)$ the previous result given in Eq. (\ref{49}), and we have defined
\bea
&&
b_{\Sg(B_s)} \equiv 
\overline{\langle \delta m^{(1)}_{\Sg(B_s)}\delta \Phi^{(1)} \rangle_{0}}
-\overline{\langle \delta m^{(1)}_{\Sg(B_s)}\rangle_{0}\langle \delta \Phi^{(1)} \rangle_{0}}\,\,\,,
\label{415}
\\ &&\nonumber  \\ &&
b_{\Da B_s} \equiv 
\overline{\langle \delta m^{(1)}_{\Da B_s}\delta \Phi^{(1)} \rangle_{0}}
-\overline{\langle \delta m^{(1)}_{\Da B_s}\rangle_{0}\langle \delta \Phi^{(1)} \rangle_{0}}\,\,\,,
\label{416}
\eea
using the measure perturbations $\delta m^{(1)}$ of Eqs. (\ref{411}) and (\ref{412}).
 
We are now in the position of comparing the different fractional corrections of Eqs. (\ref{49}), (\ref{413}) and (\ref{414}), and to discuss  their possible physical differences induced by the different embedding in the external geometry of the various averaging prescriptions.

What we need, first of all, is the explicit expression of $\delta \Phi^{(1)}$,  to be combined with $\delta m^{(1)}_{\Sg(B_s)}$ and $\delta m^{(1)}_{\Da B_s}$ in the above average integrals. Following the general results already reported in \cite{BenDayan2,BenDayan:2013gc}, and including all first order contributions but dropping, as before, the terms evaluated at the observer position, we can write $\delta \Phi^{(1)}$ as follows
\bea
\left[\delta\Phi^{(1)}\right]_{w_o, z_s}&=& 2 \kappa_s+
2\,\Xi_s\,\left[v_{\rVert\,s}+2 \int_{\eta_s}^{\eta_o}d\eta\,\pa_\eta\psi(\eta,\eta_o-\eta,\theta^a)\right]
+2 \left( 1+\Xi_s \right)\psi_s\nonumber\\
&&-\frac{4}{\eta_o-\eta_s}\int_{\eta_s}^{\eta_o}d\eta\,\psi(\eta,\eta_o-\eta,\theta^a)\,,
\label{417}
\eea
where we have defined $\Xi_s=1-\frac{1}{\Hcal_s\left( \eta_o-\eta_s \right)}$ and we have introduced the so-called lensing term $\kappa_s$, defined by
\beq
\kappa_s=\frac{1}{\eta_o-\eta_s}\int_{\eta_s}^{\eta_o}d\eta\frac{\eta-\eta_s}{\eta_o-\eta}\,\Delta_2\psi(\eta,\eta_o-\eta,\theta^a),
\label{418}
\eeq
with $\Delta_2$ the standard Laplacian operator on the unit $2$-sphere, $\Delta_2\equiv \pa^2_\theta + \cot \theta \,\pa_\theta +(\sin \theta)^{-2} \pa_\phi^2$.

In order to compute the averaged expressions (\ref{415}) and (\ref{416}) we have now to express the Bardeen potential as an integral in Fourier space over its spectral components $\psi_k(\eta)$, so that we can apply the {\it ensemble}-average conditions \cite{BenDayan:2012pp,BenDayan:2013gc} (see Appendix \ref{app:A}), 
assuming that our stochastic background of scalar perturbations is statistically homogeneous and isotropic. 
We obtain  in this way that $\langle\kappa_s\rangle_0=0$ and $\overline{\langle v_{\rVert\,s}\,\pa_rv_{\rVert\,s} \rangle_0}=0$.  
Limiting our computation to the observationally relevant range of values $0<z<5$, and including all terms which may give dominant contributions in that redshift range, we find  that we can neglect all those terms not containing at least two spacelike gradients (see Appendix A for more details on the relative importance of the various terms  induced by the perturbed geometry).  
The new geometric contributions to the fractional correction (i.e. $b_{\Sg(B_s)}$ and $b_{\Da B_s}$) can thus be analytically expressed, to leading order, as follows:
\bea
b_{\Sg(B_s)}&=&\frac{2}{\Hcal_s}\,\overline{\langle\pa_r v_{\rVert\,s}\kappa_s\rangle_0}
-2\,\Xi_s\,\overline{\langle v^2_{\rVert\,s}\rangle_0}
-2\overline{\langle v_{\rVert\,s}\kappa_s\rangle_0}
+\frac{2}{\Hcal_s}\,\Xi_s\,\overline{\langle \pa_r\psi_s\,v_{\rVert\,s} \rangle_0}
+\frac{2}{\Hcal_s}\overline{\langle \pa_r\psi_s\,\kappa_s \rangle_0}\nonumber\\
&&+\frac{2}{\Hcal_s}\left( 1+\Xi_s \right)\overline{\langle \pa_rv_{\rVert\,s}\,\psi_s \rangle_0}
-\frac{2}{\Hcal_s}\,\Xi_s\,\overline{\langle\pa_r v_{\rVert\,s} \rangle_0\langle v_{\rVert\,s}\rangle_0}
+2\,\Xi_s\,\overline{\langle v_{\rVert\,s}\rangle_0^2}\nonumber\\
&&-\frac{2}{\Hcal_s}\,\Xi_s\,\overline{\langle \pa_r\psi_s\rangle_0\langle v_{\rVert\,s} \rangle_0}\,,
\label{419}
\eea
and
\bea
b_{\Da B_s}&=&\frac{2}{\Hcal_s}\,\overline{\langle\pa_r v_{\rVert\,s}\kappa_s\rangle_0}
-2\,\Xi_s\,\left( 1-\frac{\Hcal'_s}{\Hcal^2_s} \right)\,\overline{\langle v^2_{\rVert\,s}\rangle_0}
-2\,\left( 1-\frac{\Hcal'_s}{\Hcal^2_s} \right)\overline{\langle v_{\rVert\,s}\kappa_s\rangle_0}\nonumber\\
&&+
\frac{2}{\Hcal_s}\left( 1+\Xi_s \right)\overline{\langle \pa_rv_{\rVert\,s}\,\psi_s \rangle_0}
-\frac{2}{\Hcal_s}\,\Xi_s\,\overline{\langle\pa_r v_{\rVert\,s} \rangle_0\langle v_{\rVert\,s}\rangle_0}\nonumber\\
&&+2\,\Xi_s\,\left( 1-\frac{\Hcal'_s}{\Hcal^2_s} \right)\,\overline{\langle v_{\rVert\,s}\rangle_0^2}\,.
\label{420}
\eea
See Appendix \ref{app:A} for the explicit form and a discussion of the other, non-vanishing but non-leading, first-order contributions to $b_{\Sg(B_s)}$ and $b_{\Da B_s}$ which are not explicitly included into the above equations.
In the Appendix we also provide a single compact form to express both Eqs. \eqref{419} and \eqref{420}.

All the averaged quantities appearing in the above equations are explicitly given in Appendix A in terms of integrals performed over the (dimensionless) power spectrum of scalar perturbations, $\Pcal_\Psi(k,\eta)$, defined by 
\beq
\Pcal_\Psi(k,\eta)= {k^3\over 2 \pi^2} \left|\psi_k(\eta)\right|^2 \equiv
\left[\frac{g(\eta)}{g_(\eta_o)} \right]^2 \Pcal_\Psi(k,\eta_o),
\label{421}
\eeq
where the function $g(\eta)$ controls the time evolution of the Bardeen potential as  
$\psi(\eta, x)= \left[g(\eta)/g_(\eta_o) \right] \psi_o(x)$. 

The two results (\ref{419}) and (\ref{420}) are very similar. In particular -- as anticipated in the Introduction -- 
it may be noted that the above contributions to the fractional correction of the flux, induced by two different averaging procedures, are exactly identical (at least, at the first perturbative order) provided we limit ourselves to considering the effects of redshift space distortion, i.e. to considering only the contribution of those terms  containing the average of $\pa_r v_{\rVert\,s}$.
In addition (as shown in Appendix \ref{app:A}), all the averaged contributions of Eq.  (\ref{419}) containing $\pa_r \psi_s$ (and apparently absent from Eq. (\ref{420})) can be replaced by similar contributions expressed in terms of $v_{\rVert\,s}$,  and also present in Eq. (\ref{420}).
However, as will be shown in the Appendix where we compare the numerical plots of all leading contributions, the effects of all the additional terms (besides redshift space distortion) present in Eqs. (\ref{419}) and (\ref{420})  are also non-negligible (at least in the redshift range that we are considering). 

\begin{figure}[t]
\centering
\includegraphics[scale=1.1]{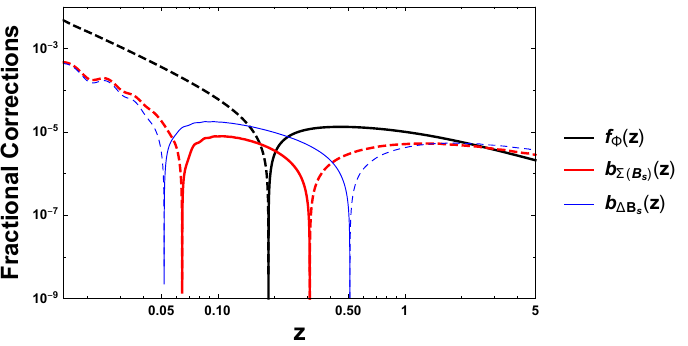}
\caption{We plot the absolute value of the fractional correction $f_\Phi$  
is compared with the absolute values of the geometric contributions $b_{\Sg(B_s)}$ and $b_{\Da B_s}$ of Eqs. (\ref{415}) and (\ref{416}), induced by the two averaging prescriptions suggested in this paper. Dashed curves correspond to negative values, solid curves to positive values.
The plots have been numerically obtained for a $\La$CDM model described by the parameters of Eqs. (\ref{422})--(\ref{425}).} 
\label{f2}
\end{figure}
\begin{figure}[h]
\centering
\includegraphics[scale=1.1]{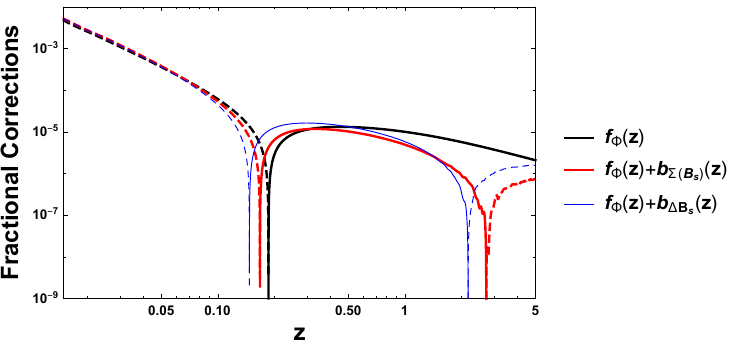}
\caption{We plot the absolute value of the three fractional corrections to the averaged flux $\overline{\langle \Phi \rangle}(z)$ defined by Eqs. (\ref{48}), (\ref{413}) and (\ref{414}), and associated, respectively, with the averaging prescriptions (\ref{311}), (\ref{310}) and (\ref{312}).
Dashed curves correspond to negative values, solid curves to positive values. The parameters of the considered $\La$CDM model are specified in  Eqs. (\ref{422})--(\ref{425}).}
\label{f3}
\end{figure}

As a consequence, there are significant differences (in both absolute value and sign, and in an appropriate range of redshift) between the two results (\ref{419}) and (\ref{420}). In order to display such differences, as well as the differences with the old result for $f_\Phi$,                 the absolute value of the old and new contributions to the fractional correction of the flux has been numerically computed and plotted as a function of $z$ in Figs. \ref{f2} and \ref{f3}. We have assumed, in particular, a model of $\La$CDM geometry with a spectrum of scalar perturbations parametrized as in Eq. (\ref{421}), where
\beq
\Pcal_\Psi(k,\eta_o)=A\left( \frac{k}{k_0} \right)^{n_s-1} {9\over 25}
\left[\frac{g(\eta_o)}{g_\infty} \right]^2T^2\left(\frac{k}{13.41\,k_\text{eq}}\right),
\label{422}
\eeq
and where $T(k)$ is the so-called transfer function which takes into account the sub-horizon evolution of modes re-entering the horizon during the radiation era.  We have expressed $T(k)$ in the Hu and Eisenstein \cite{Eisenstein:1997ik} parametrization, given by:
\bea
T(q)=\frac{L_0(q)}{L_0(q)+q^2\,C_0(q)}, ~~~~~~
L_0(q)=\log(2\,e+1.8\,q), ~~~~~~~~
C_0(q)=14.2+\frac{731}{1+62.5\,q}.
\nonumber\\ 
\label{423}
\eea
We have integrated over the spectral distribution of frequency modes using the following infraredd (IR) and ultraviolet (UV) cutoff values:
\beq
k_{\rm IR}= 3 \times 10^{-4} \,h \,{\rm Mpc}^{-1}, ~~~~~~~~
k_{\rm UV}= 0.1 \times h \,{\rm Mpc}^{-1}.
\label{423a}
\eeq
They roughly correspond to the present horizon scale and to the limiting scale of the linear spectral regime, respectively. 
Finally, we have used for the function $g(\eta)$ the standard approximated expression given in terms of the current values of the critical density parameters $\Om_{m0}$ and $\Om_\La$  (see e.g. \cite{Uzan}), namely 
\beq
g(\eta)=\frac{5}{2}\,g_\infty\frac{\Omega_m}{\Omega^{4/7}_m-\Omega_\Lambda+\left( 1+\frac{\Omega_m}{2} \right)\left( 1+\frac{\Omega_\Lambda}{70}\right)}, ~~~~~~~~~~\Omega_m=\frac{\Omega_{m0}(1+z)^3}{\Omega_{m0}(1+z)^3+\Omega_{\Lambda\,0}},
\label{424}
\eeq
where $\Omega_m+\Omega_\Lambda=\Om_{m0}+\Om_{\La 0}=1$, and where $g_\infty$ is a normalization constant fixed in such that $g(\eta_o)=1$. The numerical values of the parameters appearing in Eqs. (\ref{422}), (\ref{423}) and (\ref{424}) have been chosen, according to recent cosmological observations \cite{Ade:2015xua}, as follows:
\bea
&&
A=\,2.2\times10^{-9},~~~~~~~~~~~~n_s=0.96, ~~~~~~~~~~~~k_0=0.05\,\text{Mpc}^{-1},\nonumber\\
&&
 k_\text{eq}=0.07\,h^2\,\Omega_{m0}, ~~~~~~~~~
h=\,0.678, ~~~~~~~~~~~~ \Omega_{m0}=0.315\,.
\label{425}
\eea

As shown in particular in Fig. \ref{f3}, the differences among the three results for the fractional correction computed with the different averaging prescriptions of this paper are rather small at small redshift values (at least for the example of the flux variable that we have considered). Such differences tend to be enhanced at higher redshifts, in particular around the redshift window $2 \laq z \laq 3$, where it is clear that there are different results for the average of the flux variable. The numerical values of the fractional correction to the observed flux, however, tend to be very small ($\laq 10^{-5}-10^{-6}$) in that regime.

\section{Including non-geometric weight factors in the averaging prescription}
\label{Sec5}
\setcounter{equation}{0}

Let us finally provide an explicit example illustrating the possible role of a (non-trivial) non-geometric field $\rho (x)$, when included into the general averaging prescription according to Eqs. \eqref{21} and \eqref{23}.

We can think of such a  situation as if we were working with a generalized integral measure, $d^2 \mu \ra \rho \,d^2 \mu$. Hence, following the procedure of Sect. \ref{Sec41} and expanding $\rho$ up to first order, $\rho\simeq \rho^{(0)}\left( 1+\delta\rho^{(1)} \right)$, we simply obtain a new contribution $n_\Phi(z)$ to the fractional correction of the flux, to be linearly added to Eqs. (\ref{413}) and (\ref{414}) as
\beq
\overline{\langle \Phi/\Phi^{FLRW} \rangle}_X-1=f_\Phi(z)+b_X(z)+n_\Phi(z)\,,
\label{51}
\eeq
where we define
\beq
n_\Phi(z)\equiv
\overline{\langle\delta\rho^{(1)}\,\delta\Phi^{(1)}\rangle_0}
-\overline{\langle\delta\rho^{(1)}\rangle_0\langle\delta\Phi^{(1)}\rangle_0}\,\,\,,
\label{52}
\eeq
and we have used the symbol $X$ to denote either the averaging prescription labelled by $\Sigma(B_s)$ or the one labelled by $\Delta B_s$ . Note that the contribution of $\rho$ to the fractional correction, when computed to the lowest perturbative order, is completely independent on which type of prescription we are adopting for the geometric average.

We can now introduce a specific choice for the field $\rho(x)$. Let us adopt here for $\rho$ the density of matter sources, as in \cite{2,DiDio:2014lka}, so that for the case of averages over a given redshift bin $\Da z$ we recover the average over the number density of the sources.

The first-order contributions to the perturbations of the matter density, in the geometry described by the metric ({\ref{41}), are well known \cite{Bonvin:2011bg,Challinor:2011bk,Desjacques:2016bnm}: including all terms
(but dropping, as before, those evaluated at the observer position)} we can write\footnote{We have assumed in Eq. (\ref{53}) an evolution-bias parameter $b_{evo}=-3$, and a scale-dependent bias $b_{scale}=1$ (see e.g. \cite{Bonvin:2011bg,Challinor:2011bk,Desjacques:2016bnm}). The parameter $b_{scale}$ multiplies $\da \rho_m$, while $b_{evo}$ multiplies all the other terms of Eq. (\ref{53}). See also \cite{Desjacques:2016bnm} for the possible impact of other systematics.} 
\beq
\left[\delta\rho^{(1)}\right]_{w_o, z_s} = 3v_{\rVert\,s} + 3 \psi_s + (\delta\rho_{m})_s
+6 \int_{\eta_s}^{\eta_o}d\eta \,\pa_\eta \psi(\eta, \eta_o-\eta, \theta).
\label{53}
\eeq
On the other hand, the linear fluctuations of the matter density, $\da \rho_m$, are related as usual to the Bardeen potential $\psi$ by the Poisson-like equation, so that
\beq
(\delta\rho_{m})_s= {2\over 3}{\nabla^2 \psi_s \over \Hcal^2_s} ,
\label{54}
\eeq
where $\nabla^2$ is the standard 3-dimensional Laplacian operator. By inserting the perturbations (\ref{53}) and (\ref{417}) into the averages of Eq. (\ref{52}) we can then apply exactly the same procedure used in Sect. \ref{Sec41} to compute $b_{\Sg(B_s)}$ and $b_{\Da B_s}$. Neglecting, as before, terms without at least two spacelike gradients, as well as terms containing time derivatives and time integrals of the Bardeen potential (see the Appendix), and using the identities 
$\langle\kappa_s\rangle_0=0$,  $\overline{\langle\delta\rho_{m\,s} v_{\rVert\,s}\rangle_0}=0$, we obtain
\bea
n_\Phi(z)&=&6\,\Xi_s\,\overline{\langle v^2_{\rVert\,s}\rangle_0}
+6\,\overline{\langle v_{\rVert\,s}\kappa_s\rangle_0}
+2\,\overline{\langle\delta\rho_{m\,s}\kappa_s\rangle_0}
+2\,\left(1+\Xi_s \right)\,\overline{\langle \delta\rho_{m\,s}\,\psi_s \rangle_0}\nonumber\\
&&-6\,\Xi_s\,\overline{\langle v_{\rVert\,s}\rangle_0^2}
-2\,\Xi_s\,\overline{\langle\delta\rho_{m\,s}\rangle_0\langle v_{\rVert\,s}\rangle_0}. 
\label{55}
\eea

\begin{figure}[t]
\centering
\includegraphics[scale=1.1]{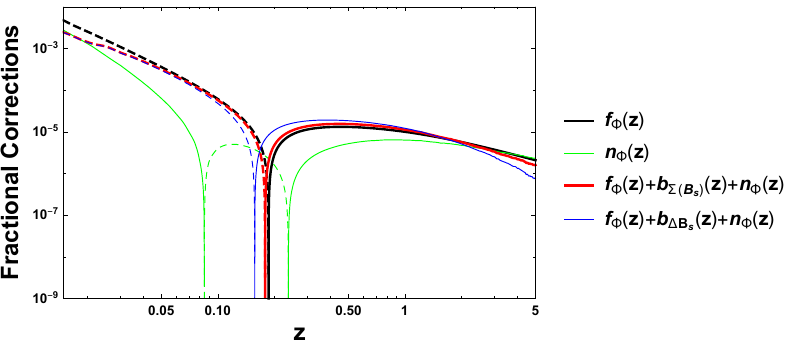}
\caption{We plot the absolute value of the fractional corrections to the flux obtained by including the matter density as a non-geometric weight factor in the average integrals. The various possible components are defined  in Eq. (\ref{51}). Dashed curves correspond to negative values, solid curves to positive values. The numerical parameters used for the plots are those specified in Eqs. (\ref{422})--(\ref{425}).}
\label{f4}
\end{figure}

The relative importance of such a contribution with respect to the contributions $f_\Phi(z)$ and $b_X(z)$, already discussed in the previous section, is illustrated in Fig. \ref{f4}. As shown by the picture when compared with Fig. \ref{f3}, including the matter density as physical weight factor in the geometric averaging prescriptions seems to have relevant effects only at large enough redshifts, $z \gaq 1$. In that regime, the presence of the weight $\rho$ seems to ``compensate" the geometric contributions of the new averages proposed in this paper, in such a way as to approach the result $f_\Phi$ computed with our original proposal of light-cone average \cite{1}. 

A similar  integral prescription for averaging the flux in the small redshift-bin limit, with the matter density $\rho$ as non geometrical weight factor, has been presented also in \cite{2} (see also the discussion of Sect. \ref{Sec3}). The numerical results, however, are different, for two reasons.
First of all we have included here the contribution of all interference terms (like the last two in Eq. \eqref{420} and the last one in Eq. \eqref{52}), which have been not taken into account in \cite{2}. Second, the matter fluctuations have been evaluated here through the Poisson equation \eqref{54}, whereas in \cite{2} they have been approximated by using a different method, which may lead to a numerical underestimation of the related effects, thus possibly explaining the differences between our results and the ones plotted in \cite{2} at higher redshifts\footnote{We thank Pierre Fleury for discussions about this point.}.

As a final remark, we emphasize  that the different prescriptions we have proposed can be tested by numerical N-body codes such as \textit{gevolution} \cite{Adamek:2015eda}. In particular, among all the possible choices of $\Scal$ as a power  of the luminosity distance $d_L$, only the average of flux (namely $d^{-2}_L$) is maximally sensitive in the redshift range $z\gaq 0.1$ to the measure adopted in the 
average (see also \cite{BenDayan:2013gc,3}). This is because the averages we have proposed contain a $\sqrt{\gamma}$ (which is nothing but   $d^2_L$) in the measure. Therefore,  the dominant lensing corrections cancel in the second-order expression of the flux \cite{BenDayan3,BenDayan:2013gc,3} making this case more sensitive to the adopted prescription. Interestingly, the recent results from \textit{gevolution} do average different power laws of the luminosity distance (see Fig. 1 of \cite{Adamek:2018rru}). Unfortunately, the plot shown there for  $d^{-2}_L$ does not look precise enough for a  precise comparison with our analytical results.

\section{Conclusion}
\label{Sec6}
\setcounter{equation}{0}

In this paper we have presented and formally defined two general and covariant averaging prescriptions, adapted to cosmological observations based on light-like signals. 

The first prescription applies to sources exactly localized on a given space-like hypersurface, and may describe a general physical situation where the flow lines along which we consider the variation of the average integral do not necessarily coincide with the world lines of the chosen observers. Also, the location of the sources does not necessarily coincide with the hypersurfaces normal to the observer world line. The second prescription applies to sources whose localization is controlled by the physical ``spread" of a given observational variable, and can  in general be confined within a thin space-time layer bounded by two very close hypersurfaces.
We have explicitly written the two different average integrals for sources located on, or between, constant redshift surfaces, and for a general cosmological geometry conveniently described by an exact metric in the GLC gauge. 

In order to illustrate the possible differences among the two types of averaging we have discussed an (important) physical example. We have computed the {\it ensemble} average of the geometric light-cone average of the received radiation flux, $\overline{\langle \Phi\rangle}$, as a function of the redshift of the emitting sources. We have adopted a simple model of $\La$CDM geometry including scalar metric perturbations to the leading, non trivial order, without anisotropic stresses. In that case, the corresponding fractional corrections (namely, the differences between the averaged results for $\overline{\langle \Phi\rangle}$ and the value $\Phi$ of the flux computed in the homogeneous FLRW background geometry) are fully controlled by the Bardeen potential $\psi$, its gradients and its time integrals. 

Including all leading contributions we have found that there are important differences already to the first perturbative order among the two averaging prescriptions, due to the different inclusion of the geometry into the integration measures. 
Such differences are analytically controlled by the factor $\Gamma_{X_s}$ (see the Appendix, in particular Eq. (\ref{compact})), which directly depends on the background geometry. From the numerical point of view it can be shown, by plotting the ratio of the results provided by the two averaging prescriptions, that such differences -- at least for the examples considered in this paper -- are of the order of ten percent or more in a large range of redshift values, while they tend to disappear in the limit of very small redshifts ($z <0.1$).

We have also numerically evaluated the possible impact of including into the average prescriptions, as a physical non-geometric weight factor, the total density $\rho$ of the matter sources. By computing again the fractional corrections of the received flux we have found that the presence of $\rho$ seems to have relevant effects (as before) only at large enough redshift values, $z \gaq 1$: in that regime, it seems to compensate the effects of the  contributions arising from the perturbations of the geometric part of the integration measure (see Fig. \ref{f4}).
Finally, we have discussed the differences between the numerical results obtained in this paper by including $\rho$ into the average integral, and previous results obtained with an equivalent averaging procedure, but applied with different  approximation methods \cite{2}.

In conclusion, we believe that the appropriate choice and the correct application of a well-posed averaging prescription is in principle of crucial importance for the correct comparison of theoretical cosmological models with increasingly precise current (and forthcoming) observational data.

\section*{Acknowledgement}
GF, MG and GM are supported in part by INFN under the program TAsP ({\it Theoretical Astroparticle Physics}). GF acknowledges support by FCT under the program {\it Stimulus} with the grant no. CEECIND/04399/2017/CP1387/CT0026. MG is supported in part by the research grant number 2017W4HA7S {\it ``NAT-NET: Neutrino and Astroparticle Theory Network"} under the program PRIN 2017 funded by the Italian Ministero dell'Universit\`a e della Ricerca (MUR). We are also grateful to Julian Adamek, Ruth Durrer and Pierre Fleury for useful discussions about the numerical results. Finally, GF, MG and GV wish to thank the hospitality and financial support of the Dipartimento di Fisica and Sezione INFN di Pisa, where an important part of this work has been carried out.

\appendix
\section{Appendix. Light-cone and {\it ensemble} averages of the flux perturbations}
\label{app:A}

Let us consider a stochastic background of metric perturbations, described by the scalar field $\psi(x)$. Assuming that the perturbations are statistically homogeneous and isotropic, $\psi$ can be decomposed in Fourier space as
\beq
\psi(\eta,\vec x)=\frac{1}{\left( 2\pi \right)^{3/2}}\int d^3 k\,E(\vec k)\psi_k(\eta)e^{i\,\vec k\cdot \vec x},
\label{a1}
\eeq
where the mode $\psi_k(\eta)$ is only dependent on $k=|\vec k|$, 
and  $E(\vec k)$ is a unit random variable  satisfying $E^*(\vec k)=E(-\vec k)$ as well as the following \textit{ensemble}-average conditions:
\beq
\overline{E(\vec k)}=0, ~~~~~~~~~~~~~~
\overline{E(\vec k)E(\vec k')}=\delta\left( \vec k+\vec k' \right).
\label{a2}
\eeq
We can then decompose the space like gradients of $\psi$, appearing in Eqs. (\ref{411}), (\ref{412}) and (\ref{417}), as follows (see also  
\cite{BenDayan:2012pp,BenDayan:2013gc}):
\bea
\pa_r\psi\left(\eta,\vec x\right)&=&\frac{1}{\left(2\pi\right)^{3/2}}\int d^3k\,E(\vec k)\psi_k\left( \eta \right)ik\cos\theta \,e^{i \vec x\cdot\vec k},\nonumber\\
\pa^2_r\psi\left(\eta,\vec x\right)&=&\frac{1}{\left(2\pi\right)^{3/2}}\int d^3k\,E(\vec k)\psi_k\left( \eta \right)\left(ik\cos\theta\right)^2 e^{i \vec x\cdot\vec k},\nonumber\\
\Delta_2\psi\left(\eta,\vec x\right)&=&-\frac{1}{\left(2\pi\right)^{3/2}}\int d^3k\,E(\vec k)\psi_k\left( \eta \right)\left( k^2r^2\,\sin^2\theta+2\,i\,kr\,\cos\theta \right)e^{i \vec x\cdot\vec k},\nonumber\\
\nabla^2\psi(\eta,\vec x)&=&-\frac{1}{(2\pi)^{3/2}}\int d^3k\,E(\vec k)\psi_k(\eta)\,k^2e^{i\vec x\cdot\vec k}\,,
\label{a3}
\eea
where we have called $\theta$ the angle between $\vec k$ and $\vec x$. 

The above derivative terms can now be inserted into the averages of Eqs. (\ref{415}) and (\ref{416}), following the same computational procedure already used in \cite{BenDayan:2012pp,BenDayan:2013gc}. Using the conditions (\ref{a2}), and noting that the unperturbed light-cone average $\langle \cdots \rangle_0$ simply corresponds, in our case, to the (normalized) angular integration over the unit homogeneous 2-sphere centered on the observer position (with measure $(\sin \theta\, d \theta d \phi)/4 \pi$), we then find that $\langle \kappa_s \rangle_0=0$ and that $\overline{\langle v_{\rVert\,s}\,\pa_rv_{\rVert\,s} \rangle_0}=0$. It turns out, in particular,  that all leading contributions appearing in Eqs. (\ref{419}) and (\ref{420}) can be expressed in terms of the following quadratic averaged expressions:
\bea
\overline{\langle v_{\rVert\,s}^2 \rangle_0}&=&\frac{1}{3}\,\left(\int_{\eta_{in}}^{\eta_s} d\eta\,\frac{a(\eta)g(\eta)}{a(\eta_s)g(\eta_o)}\right)^2\,\int \frac{dk}{k}\,k^2\,\Pcal_\Psi(k,\eta_o),\label{a4}\\
\nonumber\\
\overline{\langle v_{\rVert\,s}\,\kappa_s \rangle_0}&=&\frac{1}{2}\,
\int_{\eta_{in}}^{\eta_s}d\eta\,\int_{\eta_s}^{\eta_o}d\eta'\,\frac{a(\eta)}{a(\eta_s)}\,\frac{(\eta'-\eta_s)}{\eta_o-\eta_s}
\frac{g(\eta)g(\eta')}{g^2(\eta_o)}\nonumber\\
&&\times\int \frac{dk}{k}\,\Pcal_\Psi(k\,\eta_o)\,k^3\,\left[ (\eta_o-\eta')\,\Ical_2(k(\eta'-\eta_s))+\frac{2}{k}\,\Ical_3(k(\eta'-\eta_s)) \right],
\label{a5}\\
\nonumber\\
\overline{\langle \pa_r v_{\rVert\,s}\,\kappa_s \rangle_0}&=&-\frac{1}{2}\,\int_{\eta_s}^{\eta_o}d\eta\int_{\eta_{in}}^{\eta_s}d\eta'\,\frac{a(\eta')}{a(\eta_s)}\frac{\eta-\eta_s}{\eta_o-\eta_s}\frac{g(\eta)g(\eta')}{g^2(\eta_o)}\int\frac{dk}{k}\,k^4\,\Pcal_\Psi(k,\eta_o)\nonumber\\
&&\times\left[ \left( \eta_o-\eta \right)\,\Ical_4(k(\eta-\eta_s))+\frac{2}{k}\,\Ical_5(k(\eta-\eta_s)) \right]\,,
\label{a7}\\
\nonumber\\
\overline{\langle \pa_rv_{\rVert\,s}\rangle_0\langle v_{\rVert\,s} \rangle_0}&=&-\frac{1}{2}\,\left(\int_{\eta_{in}}^{\eta_s}d\eta\,\frac{a(\eta)g(\eta)}{a(\eta_s)g(\eta_o)}\right)^2
\int \frac{dk}{k}\,\Pcal_\Psi(k,\eta_o)k^3\Ical_3(k(\eta_o-\eta_s))\Ical_6(k(\eta_o-\eta_s)),\nonumber\\
\label{a8}\\
\overline{ \langle v_{\rVert\,s} \rangle_0^2}&=&\left(\int_{\eta_{in}}^{\eta_s}d\eta\,\frac{a(\eta)g(\eta)}{a(\eta_s)g(\eta_o)}\right)^2
\int \frac{dk}{k}\,\Pcal_\Psi(k,\eta_o)\,k^2\,\Ical^2_6(k(\eta_o-\eta_s))\,,
\label{a9}\\
\nonumber\\
\overline{\langle \pa_r v_{\rVert\,s}\,\psi_s \rangle_0}&=&-\frac{1}{3}\frac{g(\eta_s)}{g(\eta_o)}\int_{\eta_{in}}^{\eta_s} d\eta\,\frac{a(\eta)g(\eta)}{a(\eta_s)g(\eta_o)}\,\int \frac{dk}{k}\,k^2\,\Pcal_\Psi(k,\eta_o)
\label{a10}
\eea
where we have defined
\bea
&&
\Ical_2(x)=12\,\frac{\sin x}{x^4}-12\,\frac{\cos x}{x^3}-4\,\frac{\sin x}{x^2},
~~~~~~~~~~~~~~~~~~
\Ical_3(x)=4\,\frac{\sin x}{x^3}-4\,\frac{\cos x}{x^2}-2\,\frac{\sin x}{x},
\nonumber \\
&&
\Ical_5(x)=12\frac{\sin x}{x^4}-12\frac{\cos x}{x^3}-6\frac{\sin x}{x^2}+2\,\frac{\cos x}{x},
~~~~~~~
\Ical_6(x)=\frac{\sin x}{x^2}-\frac{\cos x}{x}\ ,
\nonumber\\
&&
\Ical_4(x)=48\frac{\sin x}{x^5}-48\frac{\cos x}{x^4}-20\frac{\sin x}{x^3}+4\frac{\cos x}{x^2}.
\eea
Similarly, the leading contributions appearing in Eq. (\ref{55}), and not included in the above equations, can be written explicitly as follows:
\bea
\overline{\langle \delta\rho_{m\,s}\,\psi_s \rangle_0}&=&-\frac{2}{3\,\Hcal^2_s}\left(\frac{g(\eta_s)}{g(\eta_o)}\right)^2\,\int\frac{dk}{k}k^2\,\Pcal_\Psi(k,\eta_o),
\label{a11}\\
\nonumber\\
\overline{\langle\delta\rho_{m\,s}\rangle_0\langle v_{\rVert\,s}\rangle_0}&=&\frac{2}{3\Hcal_s^2}\frac{g(\eta_s)}{g(\eta_o)}
\int_{\eta_{in}}^{\eta_s}d\eta'\frac{a(\eta')g(\eta')}{a(\eta_s)g(\eta_o)}
\int \frac{dk}{k}k^3\Pcal_\Psi(k,\eta_o)\,j_0(k(\eta_o-\eta_s))\Ical_6(k(\eta_o-\eta_s)),
\label{a12}
\nonumber\\
\\
\overline{\langle\delta\rho_{m\,s}\,\kappa_s\rangle_0}&=&\frac{4}{3\,\Hcal_s^2}
\int_{\eta_s}^{\eta_o}d\eta\frac{g(\eta_s)g(\eta)}{g^2(\eta_o)}
\int \,\frac{dk}{k}\,k^3\,\Pcal_\Psi(k,\eta_o)\,\Ical_6(k(\eta-\eta_s)), 
\label{a13}
\eea
where $j_0$ is the spherical Bessel function.

It should be noted that in the above equations we have not included terms with 
the explicit averages of $\pa_r \psi_s$ (in spite of the fact that such derivatives clearly contribute to the measure perturbations of Eq. (\ref{411}), and that they also appear among the leading terms of Eq. (\ref{419})). Interestingly enough, the reason is that  all the light-cone and {\it ensemble} averages of $\pa_r \psi_s$ can be expressed in terms of average integrals involving
$v_{\rVert\,s}$. For any operator $X_s$ we have indeed, according to our definition (\ref{46}),
\beq
\overline{\langle \pa_r\psi_s X_s \rangle_0}={\mathcal{E}_s}\,\overline{\langle v_{\rVert\,s} X_s \rangle_0}\,, ~~~~~~~~~~~~
\mathcal{E}_s \equiv \left[\int_{\eta_{in}}^{\eta_s}d\eta \frac{a(\eta)g(\eta)}{a(\eta_s)g(\eta_s)}\right]^{-1}\,.
\label{a11}
\eeq
The same occurs for terms like 
$\overline{\langle \pa_r \psi_s\rangle_0\langle X_s \rangle_0}$. 
Thanks to Eq. \eqref{a11}, Eqs. \eqref{419} and \eqref{420} can be written in identical form as follows
\bea
\!\!\!\!\!\!\!\!\!\!
b_{X_s}&=&\frac{2}{\Hcal_s}\,\overline{\langle\pa_r v_{\rVert\,s}\kappa_s\rangle_0}
-2\,\Xi_s\,\left( 1-\Gamma_{X_s} \right)\,\overline{\langle v^2_{\rVert\,s}\rangle_0}
-2\,\left( 1-\Gamma_{X_s} \right)\overline{\langle v_{\rVert\,s}\kappa_s\rangle_0}\nonumber\\
&+&
\frac{2}{\Hcal_s}\left( 1+\Xi_s \right)\overline{\langle \pa_rv_{\rVert\,s}\,\psi_s \rangle_0}
-\frac{2}{\Hcal_s}\,\Xi_s\,\overline{\langle\pa_r v_{\rVert\,s} \rangle_0\langle v_{\rVert\,s}\rangle_0}
+2\,\Xi_s\,\left( 1-\Gamma_{X_s} \right)\,\overline{\langle v_{\rVert\,s}\rangle_0^2}\,,
\label{compact}
\eea
where $X_s$ can be either $\Delta B_s$ or $\Sigma(B_s)$, and where we obtain, correspondingly, $\Gamma_{\Delta B_s}={\Hcal_s'}/{\Hcal_s^2}$  and $\Gamma_{\Sigma(B_s)}={\mathcal{E}_s}/{\Hcal_s}$.
 It may be interesting to consider the behaviour of $\Gamma_{X_s}$ at high enough redshifts, when the Universe is in the phase of matter domination with  $g=\text{constant}$ and $a\sim\eta^2$. In that regime we have $\Gamma_{\Delta B_s}=-1/2$ whereas $\Gamma_{\Sigma(B_s)}=3/2$, and we find that it is just the different value of these coefficients which almost entirely controls the different behavior of the two average prescriptions in the redshift range corresponding to matter domination. The same is true if we include the density of the matter sources in the average integrals, because its contribution is independent of the coefficient $\Gamma_{X_s}$. 
 
\begin{figure}[t]
\centering
\includegraphics[scale=1.1]{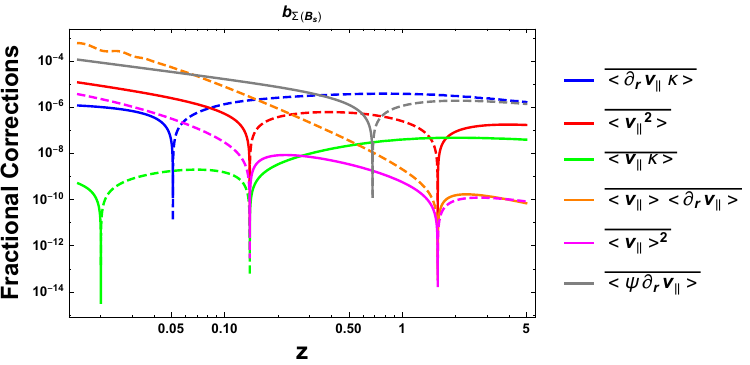}
\caption{We compare absolute value and sign of the six different types of term contributing to $b_{\Sigma(B_s)}$ as written in the form of Eq. (\ref{compact}). Each contribute is plotted by including the exact $z$-dependent coefficient controlling the relative weight of the averaged  objects with respect to the other averages. Dashed curves correspond to negative values, solid curves to positive values.}
\label{f5}
\end{figure}

\begin{figure}[t]
\centering
\includegraphics[scale=1.1]{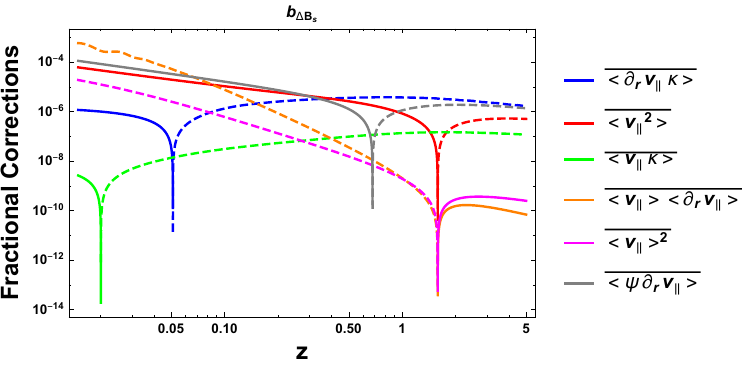}
\caption{We compare absolute value and sign of the six different types of term contributing to $b_{\Delta B_s}$ as written in Eq. (\ref{compact}). Each contribution is plotted by including the exact $z$-dependent coefficient controlling the relative weight of the averaged  objects with respect to the other averages. Dashed curves correspond to negative values, solid curves to positive values.}
\label{f6}
\end{figure}

\begin{figure}[t]
\centering
\includegraphics[scale=1.1]{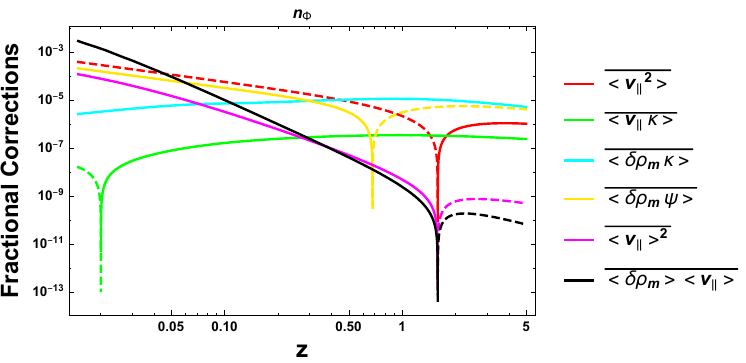}
\caption{We compare absolute value and sign of the six different types of term contributing to $n_\Phi$ (Eq. (\ref{55})). Each contribute is plotted by including the exact $z$-dependent coefficient controlling the relative weight of the averaged  objects with respect to the other averages. Dashed curves correspond to negative values, solid curves to positive values.}
\label{f7}
\end{figure}

Using Eqs. (\ref{a4})--(\ref{a11}) of this Appendix, the results for the new geometric averaged contributions to the fractional corrections of the flux can be written as in Eqs. (\ref{419}), (\ref{420}) and (\ref{55}). The single contributions of the six different types of term present in Eqs. (\ref{419}), (\ref{420}) and (\ref{55}) are explicitly illustrated (both in absolute value and sign, and for the whole redshift range $z<5$) in Figs. \ref{f5}, \ref{f6} and \ref{f7}, respectively. The sum of all contributions clearly reproduces, respectively, the behaviour of $b_{\Sg(B_s)}$ and $b_{\Da B_s}$ reported in Fig. \ref{f2}, and the behavior of $n_\Phi(z)$ reported in Fig. \ref{f4}.

We have explicitly computed also the non-leading contributions to  Eqs. (\ref{415}), (\ref{416}) and (\ref{52}), and arising, in particular, from the average of terms containing the Bardeen potential $\psi_s$, its time derivatives and its time integrals. Such terms are indeed present in the first-order perturbations of the integration measure, of the flux, and of the matter density (see  Eqs. (\ref{411}), (\ref{412}), (\ref{417}) and (\ref{52})).

Let us first consider the quadratic averages of these terms coupled to the lensing effect described by the function $\kappa_s(z)$.  Since $\langle \kappa_s \rangle_0=0$ all averages of the form $\overline{\langle \kappa_s\rangle_0\langle X_s \rangle_0}$ are vanishing (for any $X$), and we are left with the following possible contributions:
\bea
\overline{\langle\kappa_s\,\psi_s \rangle_0}&=&
-2\int_{\eta_s}^{\eta_o}d\eta\,
\frac{g(\eta)g(\eta_s)}{g^2(\eta_o)}\int \frac{dk}{k}\,k\,\Pcal_\Psi(k,\eta_o)
\,\Ical_6(k(\eta-\eta_s)), \nonumber\\
\nonumber\\
\overline{\langle\kappa_s\,\pa_\eta\psi_s \rangle_0}&=&
-2\int_{\eta_s}^{\eta_o}d\eta\,
\frac{g(\eta)g'(\eta_s)}{g^2(\eta_o)}\int \frac{dk}{k}\,k\,\Pcal_\Psi(k,\eta_o)
\,\Ical_6(k(\eta-\eta_s)), \nonumber\\
\nonumber\\
\overline{\langle\kappa_s\,\int_{\eta_s}^{\eta_o}d\eta\,\psi_s \rangle_0}&=&
-2\int_{\eta_s}^{\eta_o}d\eta\,\int_{\eta_s}^{\eta_o}\,d\eta_x\,
\frac{g(\eta)g(\eta_x)}{g^2(\eta_o)}
\frac{\left( \eta-\eta_s \right)\left( \eta_o-\eta_x \right)}{\left(\eta-\eta_x\right)\left( \eta_o-\eta_s \right)}\nonumber\\
&&\times\int \frac{dk}{k}\,k\,\Pcal_\Psi(k,\eta_o)\,\Ical_6(k(\eta-\eta_x)), \nonumber\\
\nonumber\\
\overline{\langle\kappa_s\,\int_{\eta_s}^{\eta_o}d\eta\,\pa_\eta\psi_s \rangle_0}&=&
-2\int_{\eta_s}^{\eta_o}d\eta\,\int_{\eta_s}^{\eta_o}\,d\eta_x\,
\frac{g(\eta)g'(\eta_x)}{g^2(\eta_o)}
\frac{\left( \eta-\eta_s \right)\left( \eta_o-\eta_x \right)}{\left(\eta-\eta_x\right)\left( \eta_o-\eta_s \right)}\nonumber\\
&&\times\int \frac{dk}{k}\,k\,\Pcal_\Psi(k,\eta_o)\,\Ical_6(k(\eta-\eta_x)), \nonumber\\
\nonumber\\
\overline{\langle\kappa_s\,\int_{\eta_s}^{\eta_o}d\eta\,\pa^2_\eta\psi_s \rangle_0}&=&
-2\int_{\eta_s}^{\eta_o}d\eta\,\int_{\eta_s}^{\eta_o}\,d\eta_x\,
\frac{g(\eta)g''(\eta_x)}{g^2(\eta_o)}
\frac{\left( \eta-\eta_s \right)\left( \eta_o-\eta_x \right)}{\left(\eta-\eta_x\right)\left( \eta_o-\eta_s \right)}\nonumber\\
&&\times\int \frac{dk}{k}\,k\,\Pcal_\Psi(k,\eta_o)\,\Ical_6(k(\eta-\eta_x)), 
\eea

We have numerically integrated and plotted the contributions of these terms to the fractional correction of the flux, in the redshift range $z<5$, and we have explicitly checked that (in spite of the presence of two spacelike derivatives) they are always negligible with respect to the leading contributions reported in Eqs. (\ref{419}), (\ref{420}) and (\ref{55}). In particular, the maximal amplitude of their contribution is bounded by the condition $\laq 10^{-8}$,  in the whole range of $z$  we have considered. This is not because of the coefficients controlling the relative weight of the various averaged terms, but because of the $k$-modulation of the average integrals due to the presence of the function $\Ical_6$, which is nothing but the spherical Bessel function $j_1$. 

In the same way, the quadratic averages of the Bardeen potential coupled to the ``redshift space distortion", $\pa_r v_{\rVert s}$, could produce, in addition to the leading term $\overline{\langle \pa_r v_{\rVert\,s}\,\psi_s \rangle_0}$ already included into Eqs. (\ref{419}), (\ref{420}) and (\ref{55}), also other terms  like:
\bea
\overline{\langle \pa_rv_\rVert\int_{\eta_s}^{\eta_o}d\eta\psi \rangle_0}
&=&
\frac{1}{2}\int_{\eta_s}^{\eta_o}d\eta_x\frac{g(\eta_x)}{g(\eta_o)}\int_{\eta_{in}}^{\eta_s}d\eta\frac{a(\eta)}{a(\eta_s)}\frac{g(\eta)}{g(\eta_o)}
\int \frac{dk}{k}\,k^2\,\Pcal_\Psi(k,\eta_o)\,\Ical_3\left(k\left( \eta_x-\eta_s \right)\right),
\nonumber\\
\nonumber\\
\overline{\langle \pa_rv_\rVert\,\int_{\eta_s}^{\eta_o}d\eta\,\pa_\eta\psi \rangle_0}
&=&
\frac{1}{2}\int_{\eta_s}^{\eta_o}d\eta_x\frac{g'(\eta_x)}{g(\eta_o)}
\int_{\eta_{in}}^{\eta_s}d\eta\frac{a(\eta)}{a(\eta_s)}\frac{g(\eta)}{g(\eta_o)}\int \frac{dk}{k}\,k^2\,\Pcal_\Psi(k,\eta_o)\,\Ical_3\left(k\left( \eta_x-\eta_s \right)\right).
\nonumber\\
\eea
But, as before, an explicit computation shows  that their contribution to the fractional correction of the flux  is always subleading in the range $z<5$, being suppressed by the modulation of the $k$ integrals induced by the function $\Ical_3(k)$. 

Differently from the lensing case, the contribution of terms like 
$\overline{\langle\pa_rv_{\rVert s}\rangle_0\langle X_s \rangle_0}$, where $X$ contains the potential $\psi$ and its time integrals, is not identically vaninsing. However, we have numerically checked that their amplitude is low, and their contribution to the fractional corrections is never comparable with those of the leading terms, The same is  even more  true  for all other possible quadratic averaged terms which contain less than two spatial derivatives, and that we have not even reported in this Appendix.


\end{document}